\documentclass[twocolumn,showpacs,preprintnumbers,amsmath,amssymb,groupedaddress]{revtex4}

\usepackage{graphicx}
\begin{document}
\bibliographystyle{prsty}

\title{Photoemission study of TiO$_2$/VO$_2$ interfaces}

\author{K.~Maekawa}
\affiliation{Department of Physics and Department of Complexity 
Science and Engineering, University of Tokyo, 
5-1-5 Kashiwanoha, Kashiwashi, Chiba, 277-8561, Japan}
\author{M.~Takizawa}
\affiliation{Department of Physics and Department of Complexity 
Science and Engineering, University of Tokyo, 
5-1-5 Kashiwanoha, Kashiwashi, Chiba, 277-8561, Japan}
\author{H.~Wadati}
\affiliation{Department of Physics and Department of Complexity 
Science and Engineering, University of Tokyo, 
5-1-5 Kashiwanoha, Kashiwashi, Chiba, 277-8561, Japan}
\author{T.~Yoshida}
\affiliation{Department of Physics and Department of Complexity 
Science and Engineering, University of Tokyo, 
5-1-5 Kashiwanoha, Kashiwashi, Chiba, 277-8561, Japan}
\author{A.~Fujimori}
\affiliation{Department of Physics and Department of Complexity 
Science and Engineering, University of Tokyo, 
5-1-5 Kashiwanoha, Kashiwashi, Chiba, 277-8561, Japan}
\author{H.~Kumigashira}
\affiliation{Department of Applied Chemistry, University of Tokyo, 
Bunkyo-ku, Tokyo 113-8656, Japan}
\author{M.~Oshima}
\affiliation{Department of Applied Chemistry, University of Tokyo, 
Bunkyo-ku, Tokyo 113-8656, Japan}
\author{Y.~Muraoka}
\affiliation{Institute for Solid State Physics, University of Tokyo, 5-1-5 Kashiwanoha, Kashiwashi, Chiba, 277-8581, Japan}
\author{Y.~Nagao}
\affiliation{Institute for Solid State Physics, University of Tokyo, 5-1-5 Kashiwanoha, Kashiwashi, Chiba, 277-8581, Japan}
\author{Z.~Hiroi}
\affiliation{Institute for Solid State Physics, University of Tokyo, 5-1-5 Kashiwanoha, Kashiwashi, Chiba, 277-8581, Japan}
\date{\today}

\begin{abstract}
We have measured photoemission spectra of two kinds of TiO$_2$-capped VO$_2$ thin films, 
namely, that with rutile-type TiO$_2$ (r-TiO$_2$/VO$_2$) and that with amorphous TiO$_2$ (a-TiO$_2$/VO$_2$) 
capping layers. Below the metal-insulator transition temperature of the VO$_2$ thin films, $\sim 300$ K,
metallic states were not observed for the interfaces with TiO$_2$, 
in contrast with the interfaces between the band insulator SrTiO$_3$ and the Mott insulator LaTiO$_3$
in spite of the fact that both TiO$_2$ and SrTiO$_3$ are band insulators with $d^0$ electronic configurations and 
both VO$_2$ and LaTiO$_3$ are Mott insulators with $d^1$ electronic configurations.
We discuss possible origins of this difference and suggest the importance of the polarity discontinuity of the interfaces. 
Stronger incoherent part was observed in r-TiO$_2$/VO$_2$ than in a-TiO$_2$/VO$_2$, 
suggesting Ti-V atomic diffusion due to the higher deposition temperature for r-TiO$_2$/VO$_2$.

\end{abstract}

\pacs{71.30.+h, 79.60.Dp, 73.20.-r, 73.61.-r}

\maketitle
\section{Introduction}

Strongly correlated electron systems, especially transition-metal oxides, have been the subjects of numerous studies 
because of the variety of attractive behaviors including high-temperature superconductivity,
giant magnetoresistance and metal-insulator transition (MIT) \cite{MIT}. 
Among them, Ti and V oxides are typical systems in which 
MITs between Mott-Hubbard insulators and Fermi-liquid states have been studied \cite{MIT,fujimori}.
Recently, high quality interfaces between transition-metal oxides have become available due to the development
of epitaxial thin film fabrication techniques.
Ohtomo \textit{et al}. \cite{ohtomo} have made an atomic-resolution electron-energy-loss spectroscopy study of layers of the Mott insulator LaTiO$_3$ (LTO)
embedded in the band insulator SrTiO$_3$ (STO), and found that Ti 3$d$ electrons are not completely confined within the LTO layer 
but are extended over the neighboring STO layers in spite of  the chemically abrupt interfaces.
Here, the Ti ion in LTO has the $d^1$ configuration while that in STO has the $d^0$ configuration, i.e., the empty $d$ band. 
Shibuya \textit{et al}. \cite{shibuya} reported the metallic behavior of the STO/LTO interfaces 
by measuring the electrical resistivity of STO/LTO superlattices as a function of temperature.
This metallic behavior is caused by the redistribution of electrons from
the LTO into STO layers up to several monolayers, making the electronic
structure of the interfaces analogous to that of 
the metallic La$_{1-x}$Sr$_x$TiO$_3$ alloy system \cite{tokura}.
Okamoto and Millis \cite{okamoto, okamoto2} have studied the spectral function of such systems 
by model Hartree-Fock and dynamical-mean-field-theory calculations 
and explained the novel metallic behavior at this interfaces between the two insulators, which they call ``electronic reconstruction". 
Recently, Takizawa \textit{et al}. \cite{takizawa} performed a photoemission spectroscopy study of LTO/STO interfaces using superlattice samples,
and indeed observed metallic behavior in the interface region.
In this work, we have measured photoemission spectra of another series of interfaces formed between a $d^{0}$ band insulator, TiO$_2$,
and a $d^{1}$ Mott (or Mott-Peierls) insulator, VO$_2$, in order to see 
whether the metallic behavior of the STO/LTO interfaces is a universal feature of band insulator and Mott insulator interfaces or not and to gain further insight into the mechanism which causes the metallic behavior of these interfaces.

The vanadium dioxide VO$_2$ is a well-known material which undergoes a transition 
between the high-temperature metallic phase and 
the low-temperature insulating phase \cite{morin}. 
In the high-temperature phase, the crystal has the tetragonal
rutile-type structure, while in the low-temperature phase, the V atoms
dimerize and twist, distorting 
the crystal from the tetragonal structure to a monoclinic one.
Goodenough \cite{goodenough} first discussed the MIT based on a $d_{\parallel}-d_{\pi}$ multi-band model. 
Abbate \textit{et al}. \cite{abbate} directly observed 
the splitting of the unoccupied band into the $d_{\parallel}$ and $d_{\pi}$ components 
in going from the metallic phase to the insulating phase 
by soft x-ray absorption spectroscopy.
It has been debated for many decades whether the metal-insulator transition in VO$_2$ is derived 
by electron-lattice interaction (Peierls transition) or electron-electron interaction (Mott transition).
The electronic structures of the metallic and insulating phase of VO$_2$ have been discussed emphasizing
electron-phonon interaction \cite{wentzcovitch}, electron-electron interaction \cite{rice}, and orbital degeneracy \cite{ld, tanaka}. 
So far several photoemission studies have been reported \cite{sawatzky, shin, okazaki02, shigemoto, koethe} on bulk VO$_2$ single crystals.
Recently, Muraoka \textit{et al}. prepared VO$_2$/TiO$_2$ thin films 
which undergo an MIT near room temperature \cite{muraoka, muraokaapl}.
By using thin films, it has become possible to perform detail-temperature-dependent photoemission spectroscopy (PES) studies 
utilizing the stability of the thin film surfaces \cite{okazaki, eguchi}.
In the present work, we have studied three samples; a VO$_2$ thin film capped with rutile-type TiO$_2$ (r-TiO$_2$/VO$_2$),
a VO$_2$ thin film capped with amorphous-TiO$_2$ (a-TiO$_2$/VO$_2$), and a VO$_2$ thin film without a TiO$_2$ capping layer.
We observed no metallic states at the TiO$_2$/VO$_2$ interfaces region for the low-temperature phase of VO$_2$ in contrast to the STO/LTO case \cite{takizawa}. 

\section{Experimental}
\begin{figure}
\begin{center}
\includegraphics[width=.7\linewidth]{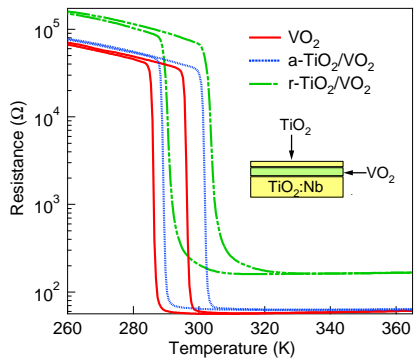}
\caption{(Color online) Electrical resistivity of the 
VO$_2$, a-TiO$_2$/VO$_2$, r-TiO$_2$/VO$_2$ thin film sample.}
\label{resistivity}
\end{center}
\end{figure}

\begin{figure}
\begin{center}
\includegraphics[width=.7\linewidth]{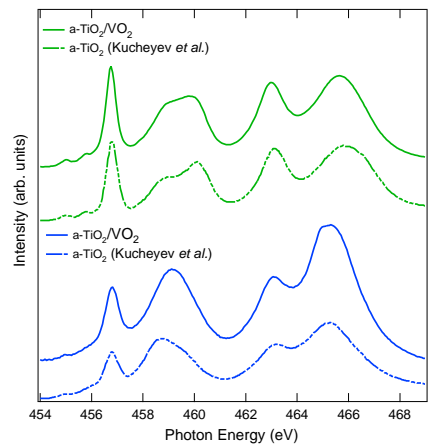}
\caption{(Color online) Ti 2$p$ XAS spectra of r-TiO$_2$/VO$_2$ 
and a-TiO$_2$/VO$_2$ compared with spectra reported 
by Kucheyev $et$ $al$. \cite{Ti2pXAS}. }
\label{Ti2pXAS}
\end{center}
\end{figure}

The films were prepared using the pulsed laser deposition (PLD) technique on Nb-doped TiO$_2$(001) substrate \cite{muraoka}.
V$_2$O$_3$ pellet was used as a target to prepare VO$_2$ thin films. During the deposition, the substrate temperature was kept at 733 K 
and the oxygen pressure was maintained at 1.0 Pa. The film thickness was about 10 -15 nm.
The r-TiO$_2$/VO$_2$ sample was prepared by depositing TiO$_2$ at 673 K on the VO$_2$ thin film.
The a-TiO$_2$/VO$_2$ sample was prepared by depositing TiO$_2$ at room temperature.
The thickness of the TiO$_2$ capping layer was one to two monolayers estimated from the deposition time.
Resistivity of these samples are shown in Fig.~\ref{resistivity}. MIT was clearly observed in all the samples.
The atomic structures of the TiO$_2$ capping layer were confirmed by comparing 
the Ti 2$p$ x-ray absorption spectra with the previous results reported by Kucheyev $et$ $al.$ \cite{Ti2pXAS} as shown in Fig.~\ref{Ti2pXAS}.
Due to the different TiO$_2$ deposition temperatures between r-TiO$_2$/VO$_2$ and a-TiO$_2$/VO$_2$,
Ti and V atoms at the r-TiO$_2$/VO$_2$ interface may be more strongly interdiffused than those at the a-TiO$_2$/VO$_2$ interface as we shall see below. 


PES measurements in the soft x-ray region were performed at BL-2C of Photon Factory, High Energy Accelerators Research Organization (KEK).
The measurements were performed under an ultrahigh vacuum of $\sim$ $10^{-10}$ Torr from room temperature down to 150 K, 
using a Scienta SES-100 electron-energy analyzer.
The total energy resolution was set to about 200 meV. 
Resonant photoemission spectroscopy (RPES) measurements at the V 2$p$ $\rightarrow$ 3$d$ 
absorption edge were performed to eliminate contributions from the
TiO$_2$ capping layers. 
The Fermi-level ($E_F$) position was determined by measuring gold spectra.
X-ray absorption spectroscopy (XAS) was performed in the total-electron-yield mode.
The samples were transferred from the PLD chamber to the spectrometers through the air, and no surface treatment was made.
During the measurements, the samples were kept at room temperature (RT) to study the metallic phase, 
or at 220 - 250 K to study the insulating phase.

\section{Results and discussion}

\begin{figure}
\begin{center}
\includegraphics[width=.7\linewidth]{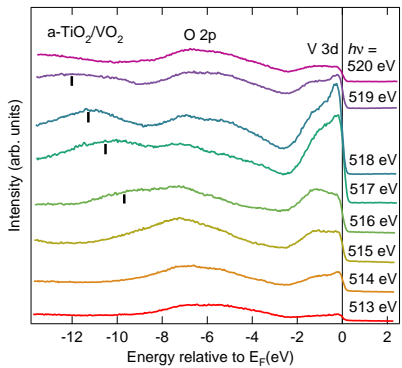}
\caption{(Color online) Valence-band V 2$p$ $\to$ 3$d$ resonant photoemission spectra of a-TiO$_2$/VO$_2$. The V 3$d$ band was strongly enhanced at 518 eV. 
Vertical lines indicate the LMM Auger peak position.}
\label{VO2RPES}
\end{center}
\end{figure}

\begin{figure}
\begin{center}
\includegraphics[width=.7\linewidth]{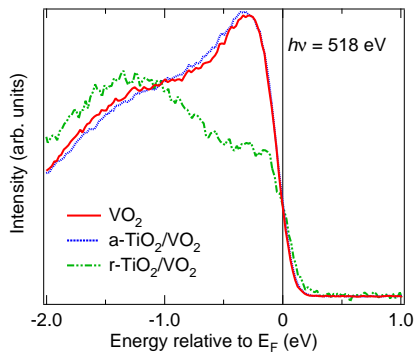}
\caption{(Color online) V 2$p$ $\to$ 3$d$ resonant photoemission spectra of VO$_2$, a-TiO$_2$/VO$_2$, and r-TiO$_2$/VO$_2$ films in the V 3$d$ band region.
The photon energy corresponds to the V 2$\textit{p}$ $\rightarrow$ 3$\textit{d}$ absorption edge and therefore, 
these spectra were free from contributions from the TiO$_2$ capping layers.}
\label{VO2metal}
\end{center}
\end{figure}

Figure \ref{VO2RPES} shows the RPES spectra of 
a-TiO$_2$/VO$_2$ for different photon energies.
The V 3$d$ band was most strongly enhanced at 518 eV, due to V 2$p$ $\to$ 3$d$ resonance.
Figure \ref{VO2metal} shows the RPES spectra of the VO$_2$, a-TiO$_2$/VO$_2$, and r-TiO$_2$/VO$_2$ 
samples in the V 3$d$ band region measured at room temperature, i.e, spectra in which VO$_2$ was metallic.
Each spectrum consisted of the coherent part (from -0.5 eV to $E_F$) and the incoherent part (from -2.0 eV to -0.5 eV). 
The spectra have been normalized to the integrated area in the V 3$d$ band region.
The spectrum of a-TiO$_2$/VO$_2$ was nearly identical to those of the VO$_2$ film.
On the other hand, those of r-TiO$_2$/VO$_2$ had stronger incoherent part than those of a-TiO$_2$/VO$_2$. 
Perhaps the amorphous TiO$_2$ layer did not form chemical bonding with the VO$_2$ surface 
and did not affect the electronic structure of the VO$_2$ film in the interface region.

\begin{figure}
\begin{center}
\includegraphics[width=.7\linewidth]{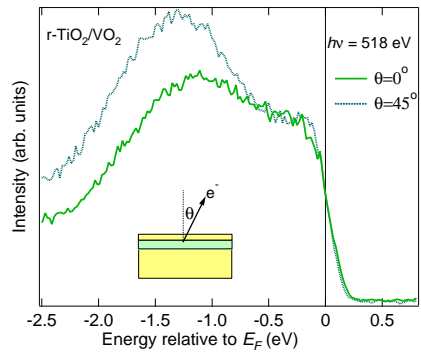}
\caption{(Color online) Angle dependence of the photoemission spectra of r-TiO$_2$/VO$_2$ in the V 3$d$ band region.
The result indicates that the strong incoherent part of r-TiO$_2$ arose from the interface region.}
\label{angdep}
\end{center}
\end{figure}

In order to see the origin of the strong incoherent part for r-TiO$_2$/VO$_2$, 
the sensitivity of the photoemission measurements to the interface was varied by changing the emission angle as shown in Fig. \ref{angdep}.
One can see that the incoherent part was enhanced as emission angle was increased, which indicates 
that the stronger incoherent part came from the interface region of r-TiO$_2$/VO$_2$. 
The different spectral weight distributions of the coherent and incoherent parts between r-TiO$_2$/VO$_2$ and a-TiO$_2$/VO$_2$ 
may have been caused by the different temperatures during the TiO$_2$-capping process.
In the r-TiO$_2$/VO$_2$ sample, for which the TiO$_2$-capping layer was deposited at a higher temperature, 
Ti and V atoms may be interdiffused to some extent, and the interface region may become somewhat like V$_{1-x}$Ti$_x$O$_2$, 
which becomes less conducting as the Ti concentration increases \cite{Tidope}.

\begin{figure}
\begin{center}
\includegraphics[width=.7\linewidth]{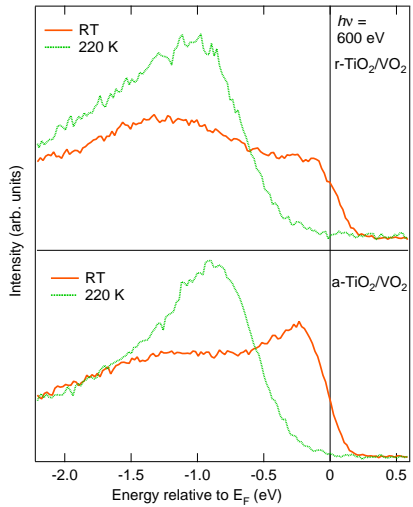}
\caption{(Color online) Photoemission spectra of r-TiO$_2$/VO$_2$ and a-TiO$_2$/VO$_2$ in the metallic and insulating phases of VO$_2$. }
\label{TiO2LT}
\end{center}
\end{figure}

The temperature dependence of the V 3$d$ valence-band spectra for r-TiO$_2$/VO$_2$ and a-TiO$_2$/VO$_2$ is shown in Fig.~\ref{TiO2LT}.
In both samples, a metallic signature, namely, the Fermi edge was not observed 
below the MIT temperature of the samples (Fig.~\ref{TiO2LT}) at the interface between the insulating TiO$_2$ and the insulating phase of VO$_2$, 
in contrast with the case of the STO/LTO interface \cite{takizawa}, although TiO$_2$ 
and VO$_2$ have $d^0$ and $d^1$ configurations, respectively, like STO and LTO.

Since the metallic behavior of the STO/LTO interfaces comes from the extended distribution of electrons from the LTO to STO layers,
the present result indicates that the electrons in VO$_2$ are confined within the VO$_2$ layer 
and are not redistributed into the TiO$_2$ capping layer.
One possible explanation for the 
confinement of electrons can be made from the view point of atoms constituting the interface: 
While TiO$_2$/VO$_2$ is composed of different transition-metal atoms, Ti and V, STO/LTO is composed of the same transition-metal atom, Ti.  
Therefore, the stronger attractive potential of the V atomic core than 
that of Ti may prevent electrons at the V site from being redistributed into the TiO$_2$ layer. 
From the crystal structure point of view, on the other hand, 
TiO$_2$/VO$_2$ has the rutile structure in which the $\textit{d}$-band width is narrower than 
that of the perovskite structure and electrons may be more easily confined across the interface.
The dielectric constant of the capping layer may also be important through the screening of electric potential: 
STO has a much larger dielectric constant than TiO$_2$ \cite{vonHippel}, 
and may weaken the attractive potential of the Mott insulating layer compared to TiO$_2$.
Moreover, the fact that VO$_2$ is not a simple Mott insulator but is a ``Mott-Peierls insulator'', 
where the Mott transition is accompanied by a crystal distortion, may also be important.
Apart from the above various possibilities, 
we point out the fact that the penetration of electrons from the LTO to STO layers can be triggered 
by the polarity discontinuity at the interface:
In order to avoid the macroscopic electric field generated across the polar LTO layer (TiO$_2$$^-$/LaO$^+$/TiO$_2$$^-$/...),
that is, in order to avoid the so-called ``polarity catastrophe" \cite{nakagawa, hotta}, part of the electronic charges should be transferred 
from the negatively charged TiO$_2$-layer side 
to the positively charged LaO-layer side of the LTO layer, 
causing the fractional occupation of the $\textit{d}$ band in both interface regions, 
and hence the metallic state.
In the case of the TiO$_2$/VO$_2$ interfaces, since the valence of V in VO$_2$ and that of Ti in TiO$_2$ are the same, such polarity discontinuity would not occur 
at the interface and the fractional occupancy of the $\textit{d}$ band would not occur. 
To confirm this scenario, systematic comparative studies of polarity discontinuous and continuous interfaces are necessary.

\section{Summary}

We have performed photoemission measurements on TiO$_2$-capped VO$_2$ thin films.
All the 3$\textit{d}$ band spectra showed the coherent and incoherent parts and the coherent part disappeared in the insulating phase of VO$_2$. 
While a-TiO$_2$/VO$_2$ showed spectra almost identical to uncapped VO$_2$ thin film, 
the incoherent part was enhanced in r-TiO$_2$/VO$_2$, which we attribute to the interdiffusion of Ti and V atoms across the interfaces.
Metallic state was not observed in the interfaces between the band insulator TiO$_2$ 
and the Mott-Peierls insulator VO$_2$ at low temperature for all the samples in contrast to the STO/LTO interfaces.
We suggest that this is originated from the absence of polarity discontinuity at the TiO$_2$/VO$_2$ interface unlike the case of STO/LTO.

\section*{Acknowledgments}
This work was supported by a Giant-in-Aid for Scientific Research (A16204024) from
JSPS and that for Priority Area ``Invention of Anomalous Quantum
Materials" (16076208) 
from the Ministry of Education, Culture, Sports, Science and 
Technology of Japan. The work at KEK-PF was done under the approval of 
Photon Factory Program Advisory Committee (Proposal No. 2005G101). 

\bibliography{VO2tex}

\end{document}